\documentclass[11pt]{article}
\usepackage[latin9]{inputenc}
\usepackage{amsfonts}
\usepackage{amsmath}
\usepackage{amssymb}
\usepackage{indentfirst}
\usepackage{graphicx}
\usepackage[colorlinks]{hyperref}
\usepackage{cite}
\usepackage{graphicx}

\numberwithin{equation}{section}
\begin{document}

\begin{titlepage}
\vspace{3cm}
\baselineskip=24pt

\begin{center}
\textbf{\LARGE{AdS Carroll Chern-Simons supergravity in 2+1 dimensions and its flat limit}}
\par\end{center}{\LARGE \par}

\begin{center}
	\vspace{1cm}
	\textbf{Lucrezia Ravera}$^{\ast}$
	\small
	\\[5mm]
	$^{\ast}$\textit{INFN, Sezione di Milano, }\\
	\textit{ Via Celoria 16, I-20133 Milano, Italy.}
	\\[5mm]
	\footnotesize
	\texttt{lucrezia.ravera@mi.infn.it}
	\par\end{center}
\vskip 20pt
\begin{abstract}
\noindent

Carroll symmetries arise when the velocity of light is sent to zero (ultra-relativistic limit).
In this paper, we present the construction of the three-dimensional Chern-Simons supergravity theory invariant under the so-called AdS Carroll superalgebra, which was obtained in the literature as a contraction of the AdS superalgebra.
The action is characterized by two coupling constants. Subsequently, we study its flat limit, obtaining the three-dimensional Chern-Simons supergravity theory invariant under the super-Carroll algebra, which is a contraction of the Poincaré superalgebra. We apply the flat limit at the level of the superalgebra, Chern-Simons action, supersymmetry transformation laws, and field equations. 

\end{abstract}
\end{titlepage}\newpage {}

\section{Introduction}

Spacetime symmetries have played a fundamental role in the understanding of diverse physical
theories such as Newtonian gravity, Maxwell's electromagnetism, special and general relativity, string and supergravity theory. Most of these models are based on relativistic
symmetries. However, in the literature models with non-relativistic symmetries have also been developed and analyzed. 

Concerning gravity theories, there are many different versions of non-relativistic gravity models (for a clear review see the recent paper \cite{Bergshoeff:2019ctr}, where the authors showed that the general method of Lie algebra expansions can be applied to reconstruct several algebras and related actions for non-relativistic gravity). These non-relativistic gravity theories are all invariant under reparametrizations, but differ in the fact that they are invariant under distinct extensions of the Galilei symmetries, the latter arising when the velocity of light is sent to infinity ($c \rightarrow \infty$, non-relativistic limit).
The simplest example is given by the so-called Galilei gravity theory, which is invariant
under the unextended Galilei symmetries \cite{Bergshoeff:2017btm}, while Newtonian gravity and
its frame-independent reformulation, Newton-Cartan gravity, are invariant under the symmetries corresponding to a central extension of the Galilei algebra, called the Bargmann algebra \cite{Bargmann, DePietri:1994je, Andringa:2010it}.

A dynamical (field theoretic) realization of Newton-Cartan geometry was formulated in \cite{Banerjee:2014nja} from the localization of the Galilean symmetry of non-relativistic matter field
theories, leading to a theory that the authors called Galilean gauge theory of gravity. Subsequently, in \cite{Banerjee:2018gqz} the authors provided an exact mapping between the aforesaid Galilian gauge theory and the Poicar\'{e} gauge theory of gravity, and applied their dictionary shewing explicitly the derivation of Newtonian gravity from Einstein's gravity.

The interest in Galilean-invariant theories with diffeomorphism invariance has increased during the years due to their relation with condensed matter systems in the context of the fractional quantum Hall effect \cite{Son:2005rv, Son:2013rqa, Geracie:2014nka} (see also \cite{Jensen:2014wha, Banerjee:2015tga} and references therein). 

Other non-relativistic theories such as non-relativistic superstrings and superbranes were studied as special points in the parameter space of M-theory \cite{Gomis:2000bd, Danielsson:2000gi}, and non-relativistic strings also attracted attention due to the fact that they appear as a possible soluble sector within string theory or M-theory \cite{Gomis:2004pw, Gomis:2005pg}.

On the other hand, there also exists another type of non-relativistic symmetry that has sporadically attracted some interest over recent years: The Carroll symmetries, which arise when the velocity of light is sent to zero ($c \rightarrow 0$, ultra-relativistic limit) \cite{LL, Bacry:1968zf}. The Carroll group introduced by L\'{e}vy-Leblond emerged as the ultra-relativistic contraction of the Poincar\'{e} group, dual to the non-relativistic contraction leading to the Galilean group. In the Carroll case, at each point of spacetime the light cone collapses to the time axis, whereas in the Galilei case it coincides with the space axis. 

Models with Carroll symmetries occurred in the study of tachyon condensation \cite{Gibbons:2002tv} and also appeared in the study of warped conformal field theories \cite{Hofman:2014loa} and in the context of tensionless strings \cite{Bagchi:2013bga, Bagchi:2015nca, Bagchi:2016yyf, Bagchi:2017cte, Bagchi:2018wsn}.

In \cite{Hartong:2015xda, Bergshoeff:2016soe} and in \cite{Bergshoeff:2017btm}, theories of Carrollian (i.e., ultra-relativistic) gravity have been developed and analyzed. 
In particular, in \cite{Bergshoeff:2016soe} the authors focused on the construction of non- and ultra-relativistic Chern-Simons (CS) type actions in $2+1$ dimensions including a spin-3 field coupled to gravity.
In \cite{Bergshoeff:2015wma}, the geometry of flat and curved (Anti-de Sitter, AdS for short) Carroll space and the symmetries of a particle moving in such a space
both in the bosonic as well as in the supersymmetric case were investigated. Afterwards, in the work \cite{Matulich:2019cdo}, which concerns the classification of gravitational theories in $2+1$ dimensions and
limits of their actions, the AdS Carroll CS gravity theory was discussed for the first time.

It has further been shown that non-relativistic symmetry groups play a remarkable role in various holography contexts \cite{Bagchi:2009my, Christensen:2013lma, Christensen:2013rfa, Hartong:2014oma, Bergshoeff:2014uea, Hartong:2015wxa, Bagchi:2010eg, Bagchi:2012cy, Bagchi:2016bcd, Lodato:2016alv, Bagchi:2019xfx, Duval:2014uva, Duval:2014lpa, Ciambelli:2018xat, Ciambelli:2018wre, Ciambelli:2018ojf, Campoleoni:2018ltl}. 
In particular, in \cite{Bagchi:2010eg} connections among the Bondi-Metzner-Sachs (BMS) algebra,\footnote{The BMS group encodes the asymptotic symmetries of asymptotically flat spacetimes along a null direction (see \cite{Bondi:1962px, Sachs:1962zza, Barnich:2009se, Ashtekar:2014zsa}).} Carrollian physics, and holography of flat space were noticed and followed up in \cite{Bagchi:2012cy} (see also \cite{Bagchi:2016bcd, Lodato:2016alv}, the latter developed in the context of supergravity, and \cite{Bagchi:2019xfx}). Moreover, in \cite{Duval:2014uva, Duval:2014lpa} conformal extensions of the Carroll group were explored and related to the BMS group, and in \cite{Ciambelli:2018xat, Ciambelli:2018wre, Ciambelli:2018ojf, Campoleoni:2018ltl} the authors shewed how Carrollian structures and geometry emerge in the framework of flat holography and fluid/gravity correspondence.

Motivated by all these interesting applications of Carroll symmetries and by the fact that a study of their supersymmetric extensions in the context of supergravity models is still lacking, in this work we present for the first time the construction of the three-dimensional CS supergravity theory invariant under the $\mathcal{N}=1$ AdS Carroll superalgebra (in $D=3$, where here and in the sequel with $D=3$ we mean $2+1$ dimensions) introduced in \cite{Bergshoeff:2015wma} (which was obtained in \cite{Bergshoeff:2015wma} as a contraction of the $\mathcal{N}=1$ AdS superalgebra) by applying the method of \cite{Concha:2016zdb}. 
Our result was also an open problem suggested in Ref. \cite{Bergshoeff:2015wma} and it represents the $\mathcal{N}=1$ supersymmetric extension of the AdS Carroll CS gravity action of \cite{Matulich:2019cdo}.

Specifically, in \cite{Concha:2016zdb} the authors presented a generalization of the standard In\"{o}n\"{u}-Wigner contraction \cite{IW, WW} by rescaling not only the generators of a Lie (super)algebra but also the arbitrary constants appearing in the components of the invariant tensor of the same Lie (super)algebra, the latter being the key ingredient for working out a CS action, invariant under the In\"{o}n\"{u}-Wigner contracted (super)algebra by construction. Thus, the development of any CS action based on an In\"{o}n\"{u}-Wigner contracted (super)algebra is assured.\footnote{Let us observe that this includes also CS theories based on degenerate, invariant bilinear forms, implying that the action is invariant under gauge transformations but it does not involve a kinematical term for each field. However, this is not the case for the AdS CS supergravity action we will present in Section \ref{S2}, which, as we will see, is indeed based on a non-degenerate, invariant bilinear form (i.e., an invariant metric) and involves a kinematical term for each field.}
In particular, the procedure presented in \cite{Concha:2016zdb} allows to obtain explicitly the CS supergravity action of a contracted superalgebra, which will be our case, since the $\mathcal{N}=1$ AdS Carroll superalgebra is obtained as a contraction of the $\mathcal{N}=1$ AdS superalgebra \cite{Bergshoeff:2015wma}.

Having constructed the three-dimensional CS supergravity theory invariant under the AdS Carroll superalgebra, we subsequently apply the flat limit ($\ell \rightarrow \infty$, being $\ell$ the length parameter) at the level of the superalgebra, CS action, gauge transformation laws, and field equations. In particular, taking the flat limit of the AdS Carroll CS supergravity action we obtain the three-dimensional CS supergravity theory invariant under the super-Carroll algebra, which is a contraction of the Poincar\'{e} superalgebra (see \cite{Bergshoeff:2015wma}).

\section{AdS Chern-Simons supergravity in 2+1 spacetime dimensions}\label{S1}

It is well assumed that a three-dimensional (super)gravity theory can be described by a CS action as a gauge theory, offering an interesting toy model to approach higher-dimensional theories \cite{DK, Deser, PvN, AT1, RPvN, Witten, AT2, NG, Howe:1995zm, Banados:1996hi, Giacomini:2006dr}. 
In the last decades, diverse three-dimensional supergravity models have been studied, and in this context there has also been a growing interest to extend AdS and Poincar\'{e} supergravity theories to
other symmetries (see \cite{Concha:2018jxx, Concha:2019icz} and references therein).

In this section, we briefly review (following \cite{Concha:2016zdb}) the construction of the $\mathfrak{osp}(2|1)\otimes \mathfrak{sp}(2)$ CS supergravity (three-dimensional AdS CS supergravity), and we pave the way for the development of the CS supergravity action in three spacetime dimensions invariant under the $D=3$ AdS Carroll superalgebra which will be presented in Section \ref{S2}.

The three-dimensional CS action is given by
\begin{equation}\label{genCS}
I_{CS}=\frac{k}{4 \pi} \int_\mathcal{M} \Big \langle A dA + \frac{2}{3} A^3 \Big \rangle \, ,
\end{equation}
where $k=1/(4G)$ is the CS level of the theory (and for gravitational theories is related to the gravitational constant $G$), $A$ corresponds to the gauge connection $1$-form, $\langle \ldots \rangle$ denotes the invariant tensor, and the integral is over a three-dimensional manifold $\mathcal{M}$.

In the case of $\mathfrak{osp}(2|1)\otimes \mathfrak{sp}(2)$, the connection $1$-form reads
\begin{equation}\label{connection}
\tilde{A} =\tilde{A}^A \tilde{T}_A = \frac{1}{2} \tilde{\omega}^{AB} \tilde{J}_{AB} + \tilde{V}^A \tilde{P}_A + \tilde{\psi}^\alpha \tilde{Q}_\alpha  \,  ,
\end{equation}
where $\tilde{T}_A = \lbrace \tilde{J}_{AB}, \tilde{P}_A , \tilde{Q}_\alpha \rbrace$ (Lorentz generators, spacetime translations, and a $2$-components Majorana spinor charge, respectively) are the $\mathfrak{osp}(2|1)\otimes \mathfrak{sp}(2)$ generators (with $A,B=0,1,2$ and $\alpha=1,2$) and $\tilde{\omega}^{AB}$, $\tilde{V}^A$, $\tilde{\psi}^\alpha$ are the spin connection, the dreibein, and the gravitino $1$-form fields, respectively.\footnote{We denote the quantities referring to $\mathfrak{osp}(2|1)\otimes \mathfrak{sp}(2)$ with a tilde symbol on the top, in order to avoid confusion with respect to the super-AdS Carroll ones, which will be introduced and treated in Section \ref{S3}.}

The corresponding curvature $2$-form $\tilde{F}= d\tilde{A} + \tilde{A} \wedge \tilde{A}$ is\footnote{In the sequel, for simplicity, we will omit the wedge product ``$\wedge$'' between differential forms. We use the metric $\eta_{AB}$ with the signature $(-,+,+)$.}
\begin{equation}\label{curv2f}
\tilde{F} = \tilde{F}^A \tilde{T}_A= \frac{1}{2} \tilde{\mathcal{R}}^{AB} \tilde{J}_{AB} + \tilde{R}^A \tilde{P}_A + \tilde{\Psi}^\alpha \tilde{Q}_\alpha \,  ,
\end{equation}
with
\begin{equation}
\begin{split}
\tilde{\mathcal{R}}^{AB} & \equiv d \tilde{\omega}^{AB} + \tilde{\omega}^A_{\phantom{A} C} \tilde{\omega}^{CB} + \frac{1}{\ell^2} \tilde{V}^A \tilde{V}^B+ \frac{1}{2\ell} \bar{\tilde{\psi}} \Gamma^{AB} \tilde{\psi} = \tilde{R}^{AB} + \frac{1}{\ell^2} \tilde{V}^A \tilde{V}^B+ \frac{1}{2\ell} \bar{\tilde{\psi}} \Gamma^{AB} \tilde{\psi}  \,  , \\
\tilde{R}^A  & \equiv d \tilde{V}^A + \tilde{\omega}^A_{\phantom{A}B} \tilde{V}^B - \frac{1}{2} \bar{\tilde{\psi}} \Gamma^A \tilde{\psi}  = \tilde{T}^A  - \frac{1}{2} \bar{\tilde{\psi}} \Gamma^A \tilde{\psi} \, , \\
\tilde{\Psi} & \equiv \tilde{\nabla} \tilde{\psi} = \tilde{D} \tilde{\psi}  + \frac{1}{2 \ell} \tilde{V}^A \Gamma_A \tilde{\psi} \,  ,
\end{split}
\end{equation}
where $\tilde{\nabla} = d + \left[ \tilde{A}, \cdot \right]$ and where $\tilde{D}=d + \tilde{\omega}$ is the Lorentz covariant derivative.

The (anti)commutation relations for $\mathfrak{osp}(2|1)\otimes \mathfrak{sp}(2)$ read
\begin{equation}\label{osp}
\begin{split}
& \left[ \tilde{J}_{AB}, \tilde{J}_{CD}\right] =\eta _{BC}\tilde{J}_{AD}-\eta _{AC}\tilde{J}_{BD}-\eta
_{BD}\tilde{J}_{AC}+\eta _{AD}\tilde{J}_{BC} \, ,  \\
& \left[ \tilde{J}_{AB},\tilde{P}_{C}\right] =\eta _{BC}\tilde{P}_{A}-\eta _{AC}\tilde{P}_{B} \,  , \quad  \quad
\left[ \tilde{P}_A , \tilde{P}_B \right] = \frac{1}{\ell^2} \tilde{J}_{AB} \,  ,   \\
& \left[ \tilde{J}_{AB},\tilde{Q}_{\alpha }\right] =-\frac{1}{2}\left( \Gamma _{AB}\tilde{Q}\right)_{\alpha } \,  , \quad \quad
\left[ \tilde{P}_{A}, \tilde{Q}_{\alpha }\right] =-\frac{1}{2 \ell} \left( \Gamma _{A}\tilde{Q}\right) _{\alpha } \,  , \\
& \left\{ \tilde{Q}_{\alpha }, \tilde{Q}_{\beta }\right\} = - \frac{1}{2 \ell} \left(\Gamma^{AB} C \right)_{\alpha \beta} \tilde{J}_{AB} + \left( \Gamma ^{A}C\right) _{\alpha \beta }\tilde{P}_{A} \,  ,
\end{split}
\end{equation}
where $\ell$ is a length parameter, $C$ denotes the charge conjugation matrix, $\Gamma_A$ represents the Dirac matrices in three dimensions (we have $C^T=-C$ and $C \Gamma^A = (C \Gamma^A)^T$), which satisfy the Clifford algebra $\lbrace \Gamma_A, \Gamma_B \rbrace = - 2 \eta_{AB}$, and $\Gamma_{AB} = \frac{1}{2} \left[\Gamma_A, \Gamma_B \right]$.

The non-vanishing components of an invariant tensor for $\mathfrak{osp}(2|1)\otimes \mathfrak{sp}(2)$ are given by
\begin{equation}
\begin{split}
& \langle \tilde{J}_{AB} \tilde{J}_{CD} \rangle = \mu_0 \left( \eta_{AD} \eta_{BC} - \eta_{AC} \eta_{BD} \right) \,  , \quad
\langle \tilde{J}_{AB} \tilde{P}_{C} \rangle  = \frac{\mu_1}{\ell} \epsilon_{ABC} \,  , \quad 
\langle \tilde{P}_{A} \tilde{P}_{B} \rangle = \frac{\mu_0}{\ell^2} \eta_{AB} \,  , \\
& \langle \tilde{Q}_\alpha \tilde{Q}_\beta \rangle = \frac{2 \left(\mu_1 -\mu_0 \right)}{\ell} C_{\alpha \beta} \,  ,
\end{split}
\end{equation}
where $\mu_0$ and $\mu_1$ are arbitrary constants and $\epsilon_{ABC}$ is the Levi-Civita symbol in three dimensions. 
For convenience, let us redefine the coefficients $\mu_0$ and $\mu_1$ as follows:\footnote{This is reminiscent of what was done in \cite{Concha:2016zdb}, since with the redefinition the coefficients appearing in the invariant tensor become dimensionful. Here, the difference is that we are also considering dimensionful generators from the very beginning, on the same lines of \cite{Concha:2018jxx}. Notice that, however, the connection $A$ in \eqref{connection} is still dimensionless, due to the fact that the spin connection, dreibein, and gravitino have dimensions of $(\text{length})^0$, $(\text{length})^1$, and $(\text{length})^{1/2}$, respectively.}
\begin{equation}
\mu_0 \rightarrow \alpha_0 \, , \quad \mu_1 	\rightarrow \ell \alpha_1 \,  .
\end{equation}
Thus, the invariant tensor takes the form
\begin{equation}\label{invtosp}
\begin{split}
& \langle \tilde{J}_{AB} \tilde{J}_{CD} \rangle = \alpha_0 \left( \eta_{AD} \eta_{BC} - \eta_{AC} \eta_{BD} \right) \, , \quad
\langle \tilde{J}_{AB} \tilde{P}_{C} \rangle = \alpha_1 \epsilon_{ABC} \,  , \quad
\langle \tilde{P}_{A} \tilde{P}_{B} \rangle = \frac{\alpha_0}{\ell^2} \eta_{AB} \,  , \\
& \langle \tilde{Q}_\alpha \tilde{Q}_\beta \rangle = 2 \left(\alpha_1 -\frac{\alpha_0}{\ell} \right) C_{\alpha \beta} \,  .
\end{split}
\end{equation}

Then, using the connection $1$-form \eqref{connection} and the invariant tensor \eqref{invtosp} in the general expression \eqref{genCS}, we obtain the $\mathfrak{osp}(2|1)\otimes \mathfrak{sp}(2)$ CS supergravity action, which reads
\begin{equation}\label{CSosp}
\begin{split}
I^{\mathfrak{osp}(2|1)\otimes \mathfrak{sp}(2)}_{CS} & = \frac{k}{4 \pi} \int_\mathcal{M} \Bigg \lbrace \frac{\alpha_0}{2} \left( \tilde{\omega}^A_{\phantom{A} B} d \tilde{\omega}^B_{\phantom{B} A} + \frac{2}{3} \tilde{\omega}^A_{\phantom{A} C} \tilde{\omega}^C_{\phantom{C}B} \tilde{\omega}^B_{\phantom{B}A} + \frac{2}{\ell^2} \tilde{V}^A \tilde{T}_A - \frac{4}{\ell} \bar{\tilde{\psi}} \tilde{\Psi} \right) \\
& + \alpha_1 \left( \epsilon_{ABC} \tilde{R}^{AB} \tilde{V}^C + \frac{1}{3 \ell^2} \epsilon_{ABC} \tilde{V}^A \tilde{V}^B \tilde{V}^C + 2 \bar{\tilde{\psi}} \tilde{\Psi} \right)  - d \left( \frac{\alpha_1}{2} \epsilon_{ABC} \tilde{\omega}^{AB} \tilde{V}^C \right) \Bigg \rbrace \,  ,
\end{split}
\end{equation}
where the term proportional to $\alpha_0$ is the exotic Lagrangian containing the so-called the Lorentz Lagrangian, a torsional part, and a contribution from the gravitino $1$-form field $\tilde{\psi}$ and its super field-strength. This action describes the most general $\mathcal{N} = 1$, $D = 3$ CS supergravity action (with cosmological constant) for the AdS supergroup \cite{Giacomini:2006dr}.

Notice that in the limit $\ell \rightarrow \infty$ the components in \eqref{invtosp} yield the non-vanishing components of the invariant tensor for the Poincar\'{e} superalgebra (the latter arising when the limit $\ell \rightarrow \infty$ is taken in \eqref{osp}), and the CS action \eqref{CSosp} reduces to
\begin{equation}\label{pcs}
\begin{split}
I^{\text{super-Poincar\'{e}}}_{CS} & = \frac{k}{4 \pi} \int_\mathcal{M} \Bigg \lbrace \frac{\alpha_0}{2} \left( \tilde{\omega}^A_{\phantom{A} B} d \tilde{\omega}^B_{\phantom{B} A} + \frac{2}{3} \tilde{\omega}^A_{\phantom{A} C} \tilde{\omega}^C_{\phantom{C}B} \tilde{\omega}^B_{\phantom{B}A} \right)  + \alpha_1 \left( \epsilon_{ABC} \tilde{R}^{AB} \tilde{V}^C + 2 \bar{\tilde{\psi}} \tilde{D}\tilde{\psi} \right) \\
& - d \left( \frac{\alpha_1}{2} \epsilon_{ABC} \tilde{\omega}^{AB} \tilde{V}^C \right) \Bigg \rbrace \,  ,
\end{split}
\end{equation}
which is the three-dimensions Poincar\'{e} CS supergravity action. Notice that the gravitino does not contribute anymore to the exotic form, which reduces to the Lorentz Lagrangian, while the term proportional to $\alpha_1$ contains the Einstein-Hilbert and Rarita-Schwinger contributions plus a boundary term.
Then, omitting the boundary term, we can see that the CS action \eqref{pcs} reproduces the pure three-dimensional supergravity action when the exotic CS term is neglected ($\alpha_0 =0$).

Now, we can pave the way for the development of a three-dimensional CS supergravity theory invariant under the AdS Carroll superalgebra (in $D=3$) introduced in \cite{Bergshoeff:2015wma}. To this aim, on the same lines of \cite{Bergshoeff:2015wma}, let us decompose the indices as
\begin{equation}\label{indexdec}
A \rightarrow (0,a) \, , \quad a=1,2 \,  .
\end{equation}
This induces the following decomposition of the generators and of the dual $1$-form fields:
\begin{equation}
\begin{split}
& \tilde{J}_{AB} \rightarrow \lbrace \tilde{J}_{ab}, \tilde{J}_{a0} \equiv \tilde{K}_a \rbrace \,  , \quad
\tilde{P}_A  \rightarrow \lbrace \tilde{P}_a , \tilde{P}_0 \equiv \tilde{H} \rbrace \,  ; \\
& \tilde{\omega}^{AB}  \rightarrow \lbrace \tilde{\omega}^{ab}, \tilde{\omega}^{a0} \equiv \tilde{k}^a \rbrace \, , \quad
\tilde{V}^A  \rightarrow \lbrace \tilde{V}^a , \tilde{V}^0 \equiv \tilde{h} \rbrace \,  .
\end{split}
\end{equation}
We also have
\begin{equation}
\Gamma_{AB} \rightarrow \lbrace \Gamma_{ab} , \Gamma_{a0} \rbrace \,  , \quad \Gamma_A \rightarrow \lbrace \Gamma_a , \Gamma_0 \rbrace \,  .
\end{equation}
Thus, the (anti)commutation relation in \eqref{osp} yield the following non-trivial ones:
\begin{equation}\label{ospdecomp}
\begin{split}
& \left[ \tilde{K}_a , \tilde{K}_b \right] = \tilde{J}_{ab} \,  , \quad \quad 
\left[ \tilde{K}_a , \tilde{J}_{bc} \right] = \delta_{ab} \tilde{K}_c - \delta_{ac}\tilde{K}_b \, , \quad \quad
\left[ \tilde{J}_{ab},\tilde{P}_{c}\right] = \delta_{bc}\tilde{P}_{a}-\delta _{ac}\tilde{P}_{b} \,  , \\
& \left[ \tilde{K}_{a},\tilde{P}_{b}\right] = - \delta_{ab} \tilde{H} \, , \quad \quad 
\left[ \tilde{K}_{a}, \tilde{H} \right] = - \tilde{P}_a \, , \quad \quad 
\left[ \tilde{P}_a , \tilde{P}_b \right] = \frac{1}{\ell^2} \tilde{J}_{ab} \, , \quad \quad 
\left[ \tilde{P}_a , \tilde{H} \right] = \frac{1}{\ell^2} \tilde{K}_{a} \, , \\
& \left[ \tilde{J}_{ab},\tilde{Q}_{\alpha }\right] = -\frac{1}{2}\left( \Gamma _{ab}\tilde{Q}\right)_{\alpha } \, , \quad \quad 
\left[ \tilde{K}_{a},\tilde{Q}_{\alpha }\right] = -\frac{1}{2}\left( \Gamma _{a0}\tilde{Q}\right)_{\alpha } \, , \\
& \left[ \tilde{P}_{a}, \tilde{Q}_{\alpha }\right] = -\frac{1}{2 \ell} \left( \Gamma _{a}\tilde{Q}\right) _{\alpha } \, , \quad \quad 
\left[ \tilde{H}, \tilde{Q}_{\alpha }\right] = -\frac{1}{2 \ell} \left( \Gamma _{0}\tilde{Q}\right) _{\alpha } \,  , \\
& \left\{ \tilde{Q}_{\alpha }, \tilde{Q}_{\beta }\right\} = - \frac{1}{2 \ell} \left(\Gamma^{ab} C \right)_{\alpha \beta} \tilde{J}_{ab} - \frac{1}{\ell} \left(\Gamma^{a0} C \right)_{\alpha \beta} \tilde{K}_{a}  + \left( \Gamma ^{a}C\right) _{\alpha \beta }\tilde{P}_{a} + \left( \Gamma ^{0}C\right) _{\alpha \beta }\tilde{H} \,  .
\end{split}
\end{equation}

Moreover, the connection $1$-form \eqref{connection} and the curvature $2$-form \eqref{curv2f} can be respectively rewritten as
\begin{equation}\label{conndec}
\tilde{A} = \frac{1}{2} \tilde{\omega}^{ab} \tilde{J}_{ab} + \tilde{k}^a \tilde{K}_a + \tilde{V}^a \tilde{P}_a + \tilde{h} \tilde{H} + \tilde{\psi}^\alpha \tilde{Q}_\alpha \,  , \quad
\tilde{F} = \frac{1}{2} \tilde{\mathcal{R}}^{ab} \tilde{J}_{ab} + \tilde{\mathcal{K}}^a \tilde{K}_a + \tilde{R}^a \tilde{P}_a + \tilde{\mathcal{H}} \tilde{H} + \tilde{\Psi}^\alpha \tilde{Q}_\alpha \,  ,
\end{equation}
with
\begin{equation}\label{curvdec}
\begin{split}
\tilde{\mathcal{R}}^{ab} & = d \tilde{\omega}^{ab} - \tilde{k}^a \tilde{k}^b + \frac{1}{\ell^2} \tilde{V}^a \tilde{V}^b + \frac{1}{2\ell} \bar{\tilde{\psi}} \Gamma^{ab} \tilde{\psi} = \tilde{R}^{ab} + \frac{1}{\ell^2} \tilde{V}^a \tilde{V}^b + \frac{1}{2\ell} \bar{\tilde{\psi}} \Gamma^{ab} \tilde{\psi} \,  , \\
\tilde{\mathcal{K}}^a  & = d \tilde{k}^a + \tilde{\omega}^a_{\phantom{a} b} \tilde{k}^b + \frac{1}{\ell^2} \tilde{V}^a \tilde{h} + \frac{1}{2 \ell} \bar{\tilde{\psi}} \Gamma^{a0} \tilde{\psi} = \tilde{\mathfrak{K}}^a  + \frac{1}{\ell^2} \tilde{V}^a \tilde{h} + \frac{1}{2 \ell} \bar{\tilde{\psi}} \Gamma^{a0} \tilde{\psi} \, , \\
\tilde{R}^a & = d \tilde{V}^a + \tilde{\omega}^a_{\phantom{a}b} \tilde{V}^b + \tilde{k}^a \tilde{h} - \frac{1}{2} \bar{\psi} \Gamma^a \tilde{\psi} = \tilde{T}^a  - \frac{1}{2} \bar{\psi} \Gamma^a \tilde{\psi} \, , \\
\tilde{\mathcal{H}} & = d \tilde{h} + \tilde{V}^a \tilde{k}_a - \frac{1}{2} \bar{\tilde{\psi}} \Gamma^0 \tilde{\psi}   = \tilde{\mathfrak{H}} - \frac{1}{2} \bar{\tilde{\psi}} \Gamma^0 \tilde{\psi}  \,  , \\
\tilde{\Psi} & = d \tilde{\psi} + \frac{1}{4} \tilde{\omega}^{ab} \Gamma_{ab} \tilde{\psi} + \frac{1}{2} \tilde{k}^{a} \Gamma_{a0} \tilde{\psi} + \frac{1}{2 \ell} \tilde{V}^a \Gamma_a \tilde{\psi}  + \frac{1}{2 \ell} \tilde{h} \Gamma_0 \tilde{\psi} \, .
\end{split}
\end{equation}

Last but not least, now the non-vanishing components of the invariant tensor takes the following form:
\begin{equation}\label{invtospdec}
\begin{split}
& \langle \tilde{J}_{ab} \tilde{J}_{cd} \rangle = \alpha_0 \left( \delta_{ad} \delta_{bc} - \delta_{ac} \delta_{bd} \right) \, , \quad 
\langle \tilde{K}_a \tilde{K}_b \rangle = - \alpha_0 \delta_{ab} \, , \\
& \langle \tilde{J}_{ab} \tilde{H} \rangle = \alpha_1 \epsilon_{ab} \, , \quad
\langle \tilde{K}_a \tilde{P}_b \rangle = - \alpha_1 \epsilon_{ab} \, , \quad
\langle \tilde{P}_{a} \tilde{P}_{b} \rangle = \frac{\alpha_0}{\ell^2} \delta_{ab} \, , \quad
\langle \tilde{H} \tilde{H} \rangle  = - \frac{\alpha_0}{\ell^2} \,  , \\
& \langle \tilde{Q}_\alpha \tilde{Q}_\beta \rangle = 2 \left(\alpha_1 -\frac{\alpha_0}{\ell} \right) C_{\alpha \beta} \, ,
\end{split}
\end{equation}
where $\epsilon_{ab} \equiv \epsilon_{0 ab} = \epsilon_{ab 0}$. As we will see in a while, these components will be fundamental in the construction of the AdS Carroll CS supergravity theory.

\section{AdS Carroll Chern-Simons supergravity in 2+1 dimensions}\label{S2}

We can now move to the construction of the three-dimensional CS supergravity theory invariant under the so-called AdS Carroll superalgebra (in $D=3$) introduced in \cite{Bergshoeff:2015wma}.

In \cite{Bergshoeff:2015wma} the $\mathcal{N}=1$ AdS Carroll superalgebra was obtained as a contraction of the $\mathcal{N}=1$ AdS superalgebra. In particular, as shown in \cite{Bergshoeff:2015wma}, the AdS Carroll superalgebra is obtained by performing the indices decomposition \eqref{indexdec} in the AdS superalgebra, which in $D=3$ yields \eqref{ospdecomp}, and subsequently, to make the Carroll contraction, by rescaling the generators with a parameter, let us call it $\sigma$, as follows:
\begin{equation}\label{resc}
\tilde{H} \rightarrow \sigma H \, , \quad \tilde{K}_a \rightarrow \sigma K_a \,  , \quad \tilde{Q}_\alpha \rightarrow \sqrt{\sigma} Q_\alpha  \, .
\end{equation}
Then, taking the limit $\sigma \rightarrow \infty$,\footnote{The limit $\sigma \rightarrow \infty$ corresponds to $\frac{1}{c} \rightarrow \infty$, being $c$ the velocity of light, that is $c \rightarrow 0$ (ultra-relativistic limit).} in $D=3$ we get the following $\mathcal{N}=1$ AdS
Carroll superalgebra generated by the set of generators $\lbrace J_{ab}, K_a, P_a, H, Q_\alpha \rbrace$ (spatial rotations, Carrollian boosts, space translations, time translations, and a two-components Majorana spinor charge, respectively)\footnote{As already said, we drop the tilde symbol on the quantities referring to the AdS Carroll superalgebra.} fulfilling the following non-trivial (anti)commutation relations:
\begin{equation}\label{adscarrollsuper}
\begin{split}
& \left[ K_a , J_{bc} \right] = \delta_{ab} K_c - \delta_{ac} K_b \, , \quad \quad
\left[ J_{ab}, P_{c}\right] =\delta_{bc}P_{a}-\delta _{ac}P_{b} \, , \quad \quad
\left[ K_{a}, P_{b}\right] = - \delta_{ab} H \, , \\
& \left[ P_a , P_b \right] = \frac{1}{\ell^2} J_{ab} \, , \quad \quad
\left[ P_a , H \right] = \frac{1}{\ell^2} K_{a} \, , \\
& \left[ J_{ab},Q_{\alpha }\right] =-\frac{1}{2}\left( \Gamma _{ab} Q \right)_{\alpha } \, , \quad \quad
\left[ P_{a}, Q_{\alpha }\right] =-\frac{1}{2 \ell} \left( \Gamma _{a}Q\right) _{\alpha } \, , \\
& \left\{ Q_{\alpha }, Q_{\beta }\right\} = - \frac{1}{\ell} \left(\Gamma^{a0} C \right)_{\alpha \beta} K_{a} + \left( \Gamma ^{0}C\right) _{\alpha \beta }H \, .
\end{split}
\end{equation}
The corresponding connection $1$-form and curvature $2$-form respectively read
\begin{equation}\label{connadscarrollsuper}
A = \frac{1}{2} \omega^{ab} J_{ab} + k^a K_a + V^a P_a + h H + \psi^\alpha Q_\alpha \,  , \quad
F = \frac{1}{2} \mathcal{R}^{ab} J_{ab} + \mathcal{K}^a K_a + R^a P_a + \mathcal{H} H + \Psi^\alpha Q_\alpha \, ,
\end{equation}
with
\begin{equation}\label{curvadscarrollsuper}
\begin{split}
\mathcal{R}^{ab} & = d \omega^{ab} + \frac{1}{\ell^2} V^a V^b = R^{ab} + \frac{1}{\ell^2} V^a V^b  \, , \\
\mathcal{K}^a  & = d k^a + \omega^a_{\phantom{a} b} k^b + \frac{1}{\ell^2} V^a h + \frac{1}{2 \ell} \bar{\psi} \Gamma^{a0} \psi = \mathfrak{K}^a  + \frac{1}{\ell^2} V^a h + \frac{1}{2 \ell} \bar{\psi} \Gamma^{a0} \psi \, , \\
R^a & = d V^a + \omega^a_{\phantom{a}b} V^b \, , \\
\mathcal{H} & = d h + V^a k_a - \frac{1}{2} \bar{\psi} \Gamma^0 \psi   = \mathfrak{H} - \frac{1}{2} \bar{\psi} \Gamma^0 \psi \,   , \\
\Psi & = d \psi + \frac{1}{4} \omega^{ab} \Gamma_{ab} \psi + \frac{1}{2 \ell} V^a \Gamma_a \psi \,  .
\end{split}
\end{equation}
Let us also mention that, considering the Bianchi identity $\nabla F =0$ ($\nabla = d + [A, \cdot]$), we obtain
\begin{equation}\label{bianchdscarrollsuper}
\begin{split}
d \mathcal{R}^{ab} & = \frac{2}{\ell^2} R^a V^b \,  , \\
D \mathcal{K}^a  & = \mathcal{R}^a_{\phantom{a} b} k^b + \frac{1}{\ell^2} R^a h - \frac{1}{\ell^2} V^a \mathcal{H} - \frac{1}{\ell} \bar{\psi} \Gamma^{a0} \Psi \, , \\
D R^a & = \mathcal{R}^a_{\phantom{a} b} V^b \,   , \\
d \mathcal{H} & = R^a k_a - V^a \mathcal{K}_a + \bar{\psi} \Gamma^0 \Psi \,  , \\
D \Psi & = \frac{1}{4} \mathcal{R}^{ab} \Gamma_{ab} \psi + \frac{1}{2 \ell} R^a \Gamma_a \psi - \frac{1}{2 \ell} V^a \Gamma_a \Psi \,  ,
\end{split}
\end{equation}
where $D=d+\omega$.

Now, in order to construct a CS action (that is an action of the form \eqref{genCS}) invariant under the super-AdS Carroll group, we require the connection $1$-form given in \eqref{connadscarrollsuper} and the corresponding non-vanishing components of the invariant tensor.

Concerning the invariant tensor, which is the fundamental ingredient for the construction of a CS action, we now apply the method of \cite{Concha:2016zdb}, which consists in rescaling not only the generators but also the coefficients appearing in the invariant tensor before applying a contraction, in order to end up with a non-trivial invariant tensor for the contracted (super)algebra on which the desired CS theory will be based.
Precisely, we consider the non-vanishing components of the invariant tensor for $\mathfrak{osp}(2|1)\otimes \mathfrak{sp}(2)$ given in \eqref{invtospdec} and rescale not only the generators in compliance with \eqref{resc} but also the coefficients appearing in \eqref{invtospdec} as follows:
\begin{equation}
\alpha_0 \rightarrow \alpha_0 \,  , \quad \alpha_1 \rightarrow \sigma \alpha_1 \, .
\end{equation}
In this way, taking the limit $\sigma \rightarrow \infty$, we end up with the following non-vanishing components of the invariant tensor for the AdS Carroll superalgebra:
\begin{equation}\label{invadscarrollsuper}
\begin{split}
& \langle J_{ab} J_{cd} \rangle = \alpha_0 \left( \delta_{ad} \delta_{bc} - \delta_{ac} \delta_{bd} \right) \, , \quad 
\langle J_{ab} H \rangle = \alpha_1 \epsilon_{ab} \, , \quad 
\langle K_a P_b \rangle = - \alpha_1 \epsilon_{ab} \, , \quad
\langle P_{a} P_{b} \rangle = \frac{\alpha_0}{\ell^2} \delta_{ab} \, , \\
& \langle Q_\alpha Q_\beta \rangle = 2 \alpha_1  C_{\alpha \beta} \,  .
\end{split}
\end{equation}

Thus, using the connection $1$-form in \eqref{connadscarrollsuper} and the non-vanishing components of the invariant tensor given in \eqref{invadscarrollsuper} in the general expression \eqref{genCS}, we can finally write the three-dimensional AdS Carroll CS supergravity action, which reads
\begin{equation}\label{CSAC}
\begin{split}
& I^{\text{super-AdS Carroll}}_{CS} = \frac{k}{4 \pi} \int_\mathcal{M} \Bigg \lbrace \frac{\alpha_0}{2} \left( \omega^a_{\phantom{a} b} R^b_{\phantom{b} a}  + \frac{2}{\ell^2} V^a R_a  \right) \\
& + \alpha_1 \left( \epsilon_{ab} R^{ab} h - 2 \epsilon_{ab} \mathfrak{K}^a V^b + \frac{1}{\ell^2} \epsilon_{ab} V^a V^b h+ 2 \bar{\psi} \Psi \right) - d \left( \frac{\alpha_1}{2} \epsilon_{ab} \omega^{ab} h - \alpha_1 \epsilon_{ab} k^a V^b \right) \Bigg \rbrace \,  ,
\end{split}
\end{equation}
written in terms of the curvatures appearing in \eqref{curvadscarrollsuper}. Notice that the term proportional to $\alpha_0$ correspond to the exotic Lagrangian involving the Lorentz contribution and a torsional piece, while it does not contain any contribution from the gravitino $1$-form field $\psi$.

The CS action \eqref{CSAC} is characterized by two coupling constants and it is invariant by construction under the super-AdS Carroll group. In particular, the local gauge transformations $\delta_\varrho A = d \varrho + \left[A, \varrho \right]$ with gauge parameter
\begin{equation}\label{gpar}
\varrho = \frac{1}{2} \varrho^{ab} J_{ab} + \kappa^a K_a + \varrho^a P_a + \tau H + \varepsilon^\alpha Q_\alpha
\end{equation}
are given by
\begin{equation}\label{gaugetr}
\begin{split}
\delta \omega^{ab} & = d \varrho^{ab} + \frac{2}{\ell^2} V^a \varrho^b \, , \\
\delta k^a & = D \kappa^a - \varrho^a_{\phantom{a} b} k^b - \frac{1}{\ell^2} \varrho^a h + \frac{1}{\ell^2} V^a \tau - \frac{1}{\ell} \bar{\varepsilon} \Gamma^{a0} \psi \, , \\
\delta V^a & = D \varrho^a - \varrho^a_{\phantom{a} b} V^b \,  , \\
\delta h & = d \tau - \varrho^a k_a + V^a \kappa_a + \bar{\varepsilon} \Gamma^0 \psi \, , \\
\delta \psi & = d \varepsilon + \frac{1}{4} \omega^{ab} \Gamma_{ab} \varepsilon - \frac{1}{4} \varrho^{ab} \Gamma_{ab} \psi - \frac{1}{2 \ell} \varrho^a \Gamma_a \psi + \frac{1}{2 \ell} V^a \Gamma_a \varepsilon \\
& = D \varepsilon - \frac{1}{4} \varrho^{ab} \Gamma_{ab} \psi - \frac{1}{2 \ell} \varrho^a \Gamma_a \psi + \frac{1}{2 \ell} V^a \Gamma_a \varepsilon \,  ,
\end{split}
\end{equation}
and, restricting ourselves to supersymmetry, we have
\begin{equation}\label{susytr}
\begin{split}
& \delta \omega^{ab} = 0 \, , \quad
\delta k^a = - \frac{1}{\ell} \bar{\varepsilon} \Gamma^{a0} \psi \, , \quad
\delta V^a = 0 \, , \quad
\delta h = \bar{\varepsilon} \Gamma^0 \psi \, , \\
& \delta \psi = d \varepsilon + \frac{1}{4} \omega^{ab} \Gamma_{ab} \varepsilon + \frac{1}{2 \ell} V^a \Gamma_a \varepsilon = D \varepsilon + \frac{1}{2 \ell} V^a \Gamma_a \varepsilon \, .
\end{split}
\end{equation}
Let us also mention, for the sake of completeness, that from the gauge variation of the curvature $2$-form $F$ in \eqref{connadscarrollsuper}, $\delta_\varrho F = \left[ F, \varrho \right] $, we can write the gauge transformations
\begin{equation}\label{curvtr}
\begin{split}
\delta \mathcal{R}^{ab} & = \frac{2}{\ell^2} R^a \varrho^b \,  , \\
\delta \mathcal{K}^a  & = \mathcal{R}^a_{\phantom{a} b} \kappa^b - \varrho^a_{\phantom{a} b} \mathcal{K}^b - \frac{1}{\ell^2} \varrho^a \mathcal{H} + \frac{1}{\ell^2} R^a \tau - \frac{1}{\ell} \bar{\varepsilon} \Gamma^{a0} \Psi \,  , \\
\delta R^a & = \mathcal{R}^a_{\phantom{a} b} \varrho^b - \varrho^a_{\phantom{a} b} R^b \,  , \\
\delta \mathcal{H} & = - \varrho^a \mathcal{K}_a + R^a \kappa_a + \bar{\varepsilon} \Gamma^0 \Psi \, , \\
\delta \Psi & = \frac{1}{4} \mathcal{R}^{ab} \Gamma_{ab} \varepsilon - \frac{1}{4} \varrho^{ab} \Gamma_{ab} \Psi - \frac{1}{2 \ell} \varrho^a \Gamma_a \Psi + \frac{1}{2 \ell} R^a \Gamma_a \varepsilon \, .
\end{split}
\end{equation}

The equations of motion derived from the variation of the action \eqref{CSAC} with respect to the fields $\omega^{ab}$, $k^a$, $V^a$, $h$, and $\psi$ are
\begin{equation}\label{eom}
\begin{split}
& \delta \omega^{ab} \; : \quad \alpha_0 \mathcal{R}^{ab} + \alpha_1 \epsilon^{ab} \mathcal{H}= 0 \, , \quad \quad \quad
\delta k^a \; : \quad  \alpha_1 R^a = 0 \, , \\
& \delta V^a \; : \quad \frac{\alpha_0}{\ell^2} R^a + 2 \alpha_1 \epsilon_{ab} \mathcal{K}^b = 0 \,  , \quad \quad \quad 
\delta h \; : \quad \alpha_1 \mathcal{R}^{ab} = 0  \, , \\
& \delta \psi \; : \quad \quad \alpha_1 \Psi = 0 \,  ,
\end{split}
\end{equation}
respectively.\footnote{In these calculations we have also exploited the identities $\epsilon_{ABC} \epsilon^{ADE} = - \left( \delta_B^D \delta_C^E - \delta_B^E \delta_C^D \right)$, $\epsilon_{ABC} \epsilon^{ABD} = - 2 \delta_C^D$, that is, in particular, $\epsilon_{ab} \epsilon^{ac} = - \delta_b^{c}$, and $\Gamma_{AB} = - \epsilon_{ABC}\Gamma^C$, which implies $\Gamma_{ab} = -\epsilon_{ab} \Gamma^0$, $\Gamma_{a0} = \epsilon_{ab} \Gamma^b$.}
In particular, when $\alpha_1 \neq 0$, the equations of motion in \eqref{eom} reduce to the vanishing of the super-AdS Carroll curvature $2$-forms, namely
\begin{equation}\label{eomvac}
\mathcal{R}^{ab} = 0 \,  , \quad \mathcal{K}^a = 0 \, , \quad R^a = 0 \, , \quad \mathcal{H}=0 \, , \quad \Psi =0 \, .
\end{equation}
Let us observe that $\alpha_1 \neq 0$ is a sufficient condition to recover \eqref{eomvac}, meaning that one could consistently set $\alpha_0=0$, which corresponds to the vanishing of the exotic term in the CS action \eqref{CSAC}.

Notice also that if we restrict ourselves to the purely bosonic part of the action \eqref{CSAC} (that is if we neglect supersymmetry), we get
\begin{equation}\label{CSACbos}
\begin{split}
I^{\text{AdS Carroll}}_{CS} & = \frac{k}{4 \pi} \int_\mathcal{M} \Bigg \lbrace \frac{\alpha_0}{2} \left( \omega^a_{\phantom{a} b} R^b_{\phantom{b} a}  + \frac{2}{\ell^2} V^a R_a  \right) + \alpha_1 \left( \epsilon_{ab} R^{ab} h - 2 \epsilon_{ab} \mathfrak{K}^a V^b + \frac{1}{\ell^2} \epsilon_{ab} V^a V^b h \right) \\
& - d \left( \frac{\alpha_1}{2} \epsilon_{ab} \omega^{ab} h - \alpha_1 \epsilon_{ab} k^a V^b \right) \Bigg \rbrace \, ,
\end{split}
\end{equation}
which is the three-dimensional CS gravity action invariant under the $D=3$ AdS Carroll algebra \cite{Matulich:2019cdo} (see also \cite{Bergshoeff:2015wma})
\begin{equation}\label{adscarroll}
\begin{split}
& \left[ K_a , J_{bc} \right] = \delta_{ab} K_c - \delta_{ac} K_b \, , \quad \quad
\left[ J_{ab}, P_{c}\right] =\delta_{bc}P_{a}-\delta _{ac}P_{b} \, , \quad \quad
\left[ K_{a}, P_{b}\right] = - \delta_{ab} H \, , \\
& \left[ P_a , P_b \right] = \frac{1}{\ell^2} J_{ab} \, , \quad \quad
\left[ P_a , H \right] = \frac{1}{\ell^2} K_{a} \, , 
\end{split}
\end{equation}
bosonic subalgebra of the AdS-Carroll superalgebra \eqref{adscarrollsuper}.

\section{Study of the flat limit}\label{S3}

In the sequel, we finally study the flat limit ($\ell \rightarrow \infty$), which can be directly applied to the AdS Carroll superalgebra \eqref{adscarrollsuper}, to the curvature $2$-forms \eqref{curvadscarrollsuper}, to the CS action \eqref{CSAC}, to the transformation laws \eqref{gaugetr}, \eqref{susytr}, and \eqref{curvtr}, and to the field equations \eqref{eom}.

In the limit $\ell \rightarrow \infty$, the (anti)commutation relations of the superalgebra \eqref{adscarrollsuper} reduce to the following non-trivial ones:
\begin{equation}\label{adscarrollsuperF}
\begin{split}
& \left[ K_a , J_{bc} \right] = \delta_{ab} K_c - \delta_{ac} K_b \,  , \quad \quad
\left[ J_{ab}, P_{c}\right] =\delta_{bc}P_{a}-\delta _{ac}P_{b} \, , \\
& \left[ K_{a}, P_{b}\right] = - \delta_{ab} H \, , \quad \quad 
\left[ P_a , H \right]  = \frac{1}{\ell^2} K_{a} \, , \\
& \left[ J_{ab},Q_{\alpha }\right] =-\frac{1}{2}\left( \Gamma _{ab} Q \right)_{\alpha } \, , \quad \quad
\left\{ Q_{\alpha }, Q_{\beta }\right\} = \left( \Gamma ^{0}C\right) _{\alpha \beta }H \,  .
\end{split}
\end{equation}
These are the (anti)commutation relations of the $\mathcal{N}=1$, $D=3$ super-Carroll algebra, which can also be obtained as a contraction of the $D=3$ super-Poincar\'{e} algebra (see \cite{Bergshoeff:2015wma} and references therein).

As $\ell \rightarrow \infty$, the $2$-form curvatures \eqref{curvadscarrollsuper} and the corresponding Bianchi identities \eqref{bianchdscarrollsuper} become
\begin{equation}\label{curvadscarrollsuperF}
\begin{aligned}[c]
& \mathcal{R}^{ab} = d \omega^{ab} \, , \\
& \mathcal{K}^a = d k^a + \omega^a_{\phantom{a} b} k^b \, , \\
& R^a = d V^a + \omega^a_{\phantom{a}b} V^b \, , \\
& \mathcal{H} = d h + V^a k_a - \tfrac{1}{2} \bar{\psi} \Gamma^0 \psi   = \mathfrak{H} - \tfrac{1}{2} \bar{\psi} \Gamma^0 \psi  \,  , \\
& \Psi = d \psi + \tfrac{1}{4} \omega^{ab} \Gamma_{ab} \psi \, ,
\end{aligned}
\quad \quad
\begin{aligned}[c]
& d \mathcal{R}^{ab} = 0 \,  , \\
& D \mathcal{K}^a = \mathcal{R}^a_{\phantom{a} b} k^b \,   , \\
& D R^a = \mathcal{R}^a_{\phantom{a} b} V^b \,  , \\
& d \mathcal{H} = R^a k_a - V^a \mathcal{K}_a + \bar{\psi} \Gamma^0 \Psi \, , \\
& D \Psi = \tfrac{1}{4} \mathcal{R}^{ab} \Gamma_{ab} \psi \,  .
\end{aligned}
\end{equation}
We can also take the $\ell \rightarrow \infty$ limit of the invariant tensor \eqref{invadscarrollsuper}, which yields
\begin{equation}\label{invadscarrollsuperF}
\begin{split}
& \langle J_{ab} J_{cd} \rangle = \alpha_0 \left( \delta_{ad} \delta_{bc} - \delta_{ac} \delta_{bd} \right) \, , \quad 
\langle J_{ab} H \rangle = \alpha_1 \epsilon_{ab} \, , \quad 
\langle K_a P_b \rangle = - \alpha_1 \epsilon_{ab} \, , \quad
\langle P_a P_b \rangle = 0 \, , \\
& \langle Q_\alpha Q_\beta \rangle = 2 \alpha_1  C_{\alpha \beta} \, .
\end{split}
\end{equation}
Then, applying the $\ell \rightarrow \infty$ limit to the CS action \eqref{CSAC} we get 
\begin{equation}\label{CSC}
\begin{split}
I^{\text{super-Carroll}}_{CS} & = \frac{k}{4 \pi} \int_\mathcal{M} \Bigg \lbrace \frac{\alpha_0}{2} \left( \omega^a_{\phantom{a} b} \mathcal{R}^b_{\phantom{b} a} \right) + \alpha_1 \left( \epsilon_{ab} \mathcal{R}^{ab} h - 2 \epsilon_{ab} \mathcal{K}^a V^b + 2 \bar{\psi} \Psi \right)  \\
& - d \left( \frac{\alpha_1}{2} \epsilon_{ab} \omega^{ab} h - \alpha_1 \epsilon_{ab} k^a V^b \right) \Bigg \rbrace \, ,
\end{split}
\end{equation}
which is now written in terms of the super-Carroll curvatures appearing in \eqref{curvadscarrollsuperF} (they should not be confused with the super-AdS Carroll ones).
Observe that the same action can be derived by using the non-vanishing components in \eqref{invadscarrollsuperF} and the connection $1$-form for the Carroll superalgebra \eqref{adscarrollsuperF} in the general expression \eqref{genCS}. Furthermore, we can see that the exotic term, proportional to $\alpha_0$, now reduces purely to the so-called Lorentz Lagrangian, without any contribution from the torsion or the $1$-form field $\psi$.

The CS action \eqref{CSC} is invariant by construction under the super-Carroll group. Thus, by taking the flat limit of the CS supergravity action \eqref{CSAC}, which is invariant under the super-AdS Carroll group, we have obtained the three-dimensional CS supergravity theory invariant under the super-Carroll algebra. 

In particular, concerning the flat limit of the gauge transformations \eqref{gaugetr}, we have
\begin{equation}\label{gaugetrF}
\begin{split}
\delta \omega^{ab} & = d \varrho^{ab} \, , \\
\delta k^a & = D \kappa^a - \varrho^a_{\phantom{a} b} k^b \, , \\
\delta V^a & = D \varrho^a - \varrho^a_{\phantom{a} b} V^b \, , \\
\delta h & = d \tau - \varrho^a k_a + V^a \kappa_a + \bar{\varepsilon} \Gamma^0 \psi \, , \\
\delta \psi & = d \varepsilon + \frac{1}{4} \omega^{ab} \Gamma_{ab} \varepsilon - \frac{1}{4} \varrho^{ab} \Gamma_{ab} \psi = D \varepsilon - \frac{1}{4} \varrho^{ab} \Gamma_{ab} \psi \, ,
\end{split}
\end{equation}
which are the super-Carroll local gauge transformations $\delta_\varrho A = d \varrho + \left[A, \varrho \right]$, where, with a little abuse of notation with respect to Section \ref{S2}, we have now denoted by $A$ the super-Carroll $1$-form connection and where the gauge parameter is given by \eqref{gpar} ($J_{ab}$, $K_a$, $P_a$, $H$, and $Q_\alpha$ are now the generators of the Carroll superalgebra \eqref{adscarrollsuperF}).
The restriction to supersymmetry transformations reads
\begin{equation}\label{susytrF}
\delta \omega^{ab} = 0 \, , \quad 
\delta k^a = 0 \, , \quad
\delta V^a = 0 \, , \quad
\delta h = \bar{\varepsilon} \Gamma^0 \psi \, , \quad
\delta \psi = d \varepsilon + \frac{1}{4} \omega^{ab} \Gamma_{ab} \varepsilon = D \varepsilon \, .
\end{equation}
Furthermore, the gauge transformations \eqref{curvtr} in the limit $\ell \rightarrow \infty$ become
\begin{equation}\label{curvtrF}
\begin{split}
\delta \mathcal{R}^{ab} & = 0 \, , \\
\delta \mathcal{K}^a  & = \mathcal{R}^a_{\phantom{a} b} \kappa^b - \varrho^a_{\phantom{a} b} \mathcal{K}^b \, , \\
\delta R^a & = \mathcal{R}^a_{\phantom{a} b} \varrho^b - \varrho^a_{\phantom{a} b} R^b \, , \\
\delta \mathcal{H} & = - \varrho^a \mathcal{K}_a + R^a \kappa_a + \bar{\varepsilon} \Gamma^0 \Psi \, , \\
\delta \Psi & = \frac{1}{4} \mathcal{R}^{ab} \Gamma_{ab} \varepsilon - \frac{1}{4} \varrho^{ab} \Gamma_{ab} \Psi \, ,
\end{split}
\end{equation}
which are now the transformations for the super-Carroll curvature $2$-forms in \eqref{curvadscarrollsuperF}.

Finally, the equations of motion derived from the action \eqref{CSC} (flat limit of the equations \eqref{eom}) are
\begin{equation}\label{eomF}
\begin{split}
& \delta \omega^{ab} \; : \quad \alpha_0 \mathcal{R}^{ab} + \alpha_1 \epsilon^{ab} \mathcal{H}= 0 \, , \quad \quad \quad
\delta k^a \; : \quad \alpha_1 R^a = 0 \, , \\
& \delta V^a \; : \quad 2 \alpha_1 \epsilon_{ab} \mathcal{K}^b = 0  \, , \quad \quad \quad \quad \quad \quad 
\delta h \; : \quad \alpha_1 \mathcal{R}^{ab} = 0 \,  , \\
& \delta \psi \; : \quad \alpha_1 \Psi = 0 \,  .
\end{split}
\end{equation}
When $\alpha_1 \neq 0$, the equations of motion in \eqref{eomF} reduce to the vanishing of the super-Carroll curvature $2$-forms given in \eqref{curvadscarrollsuperF}.
Let us observe that, analogously to what happened in Section \ref{S3} in the case of the super-AdS Carroll CS supergravity theory, $\alpha_1 \neq 0$ is a sufficient condition to recover the vanishing of the super-Carroll curvature $2$-forms, meaning that one could consistently set $\alpha_0=0$, which corresponds to the vanishing of the exotic term (Lorentz Lagrangian) in the CS action \eqref{CSC}.

We conclude noting that the restriction to the purely bosonic part of the action \eqref{CSC} yields
\begin{equation}\label{CSCbos}
I^{\text{Carroll}}_{CS} = \frac{k}{4 \pi} \int_\mathcal{M} \Bigg \lbrace \frac{\alpha_0}{2} \left( \omega^a_{\phantom{a} b} \mathcal{R}^b_{\phantom{b} a} \right) + \alpha_1 \left( \epsilon_{ab} \mathcal{R}^{ab} h - 2 \epsilon_{ab} \mathcal{K}^a V^b \right) - d \left( \frac{\alpha_1}{2} \epsilon_{ab} \omega^{ab} h - \alpha_1 \epsilon_{ab} k^a V^b \right) \Bigg \rbrace \, ,
\end{equation}
which is the three-dimensional CS gravity action invariant under the $D=3$ Carroll algebra \cite{LL, Bacry:1968zf} (bosonic subalgebra of the Carroll superalgebra \eqref{adscarrollsuperF}), whose non-trivial commutation relations read
\begin{equation}\label{carroll}
\left[ K_a , J_{bc} \right] = \delta_{ab} K_c - \delta_{ac} K_b \,  , \quad \quad 
\left[ J_{ab}, P_{c}\right] =\delta_{bc}P_{a}-\delta _{ac}P_{b} \, , \quad \quad
\left[ K_{a}, P_{b}\right] = - \delta_{ab} H \,  .
\end{equation}
The CS action \eqref{CSCbos}, as argued in \cite{Bergshoeff:2017btm}, is equivalent to the action found in \cite{Bergshoeff:2017btm} if we take the $D=3$ case in the same paper.

\section{Final remarks}\label{finalremarks}

Motivated by the recent development of applications of Carroll symmetries (in particular, by their prominent role in the context of holography), and by the fact that, nevertheless, the study of their supersymmetric extensions in the context of supergravity models was still unexplored, in this paper we have presented, for the first time in the literature, the three-dimensional CS supergravity theory invariant under the $\mathcal{N}=1$ AdS Carroll superalgebra introduced in \cite{Bergshoeff:2015wma} (our result was also an open problem suggested in Ref. \cite{Bergshoeff:2015wma} and it represents the $\mathcal{N}=1$ supersymmetric extension of the AdS Carroll CS gravity action of \cite{Matulich:2019cdo}).
We have obtained the three-dimensional AdS Carroll CS supergravity action by applying the method of \cite{Concha:2016zdb}.
The aforesaid action is written in \eqref{CSAC}, and it is characterized by two coupling constants.
The restriction to the purely bosonic part of the three-dimensional CS supergravity theory invariant under the $\mathcal{N}=1$ super-AdS Carroll algebra yields the three-dimensional CS gravity action invariant under the AdS Carroll algebra \cite{Matulich:2019cdo}, bosonic subalgebra of the AdS Carroll superalgebra.

Subsequently, we have applied the flat limit ($\ell \rightarrow \infty$) at the level of the superalgebra, CS action, supersymmetry transformation laws, and field equations.
The limit $\ell \rightarrow \infty$ of the $\mathcal{N}=1$ AdS Carroll superalgebra yields the $\mathcal{N}=1$ super-Carroll algebra and, in particular, taking the flat limit of the AdS Carroll CS supergravity action we have obtained the three-dimensional (flat) CS supergravity theory invariant under the super-Carroll algebra. Restricting ourselves to the purely bosonic part of the super-Carroll CS supergravity action, we have obtained the three-dimensional CS gravity action invariant under the Carroll algebra \cite{LL, Bacry:1968zf}, bosonic subalgebra of the Carroll superalgebra. This CS action is equivalent to the action found in \cite{Bergshoeff:2017btm} if we consider the $D=3$ case in the same paper.

This work could shed some light on the development of supersymmetric extensions of \cite{Bergshoeff:2016soe}, and might also represents a starting point to go further in the analysis of supersymmetry invariance of flat supergravity in the presence of a non-trivial boundary, along the lines of \cite{Concha:2018ywv}. Furthermore, having well defined three-dimensional CS (super)gravity theories respectively invariant under the AdS-Carroll and Carroll (super)algebras, it would be intriguing to go beyond and study the asymptotic symmetry of these theories (following, for example, the prescription of \cite{Concha:2018zeb}).
Another future work could consist in considering the $\mathcal{N}$-extended AdS Carroll superalgebras in order to build $(p,q)$-type CS supergravity models (on the same lines of what was done in \cite{Concha:2019icz} in the context of Maxwell supersymmetries) by adopting the same prescription followed in the present paper. Some work is in progress on these points.

\section{Acknowledgments}

The author wishes to thank Laura Andrianopoli for the inspiring discussions that have aroused her interest in Carroll symmetries. The author also acknowledges enlightening dialogues with Stefan Prohazka.

\end{document}